\documentclass[prl,twocolumn,showpacs,preprintnumbers,amsmath,amssymb]{revtex4}

\usepackage{epsfig}
\usepackage{amsmath,amssymb}
\usepackage{graphicx}
\usepackage{dcolumn}
\usepackage{bm}
\newcommand{\unit}[1]{\ensuremath{\,\mathrm{#1}}}
\newcommand{\Yb}{\ensuremath{^{171}\mathrm{Yb}^+}}

\newcommand{\MHz}{\unit{MHz}}
\newcommand{\ket}[1]{\ensuremath{\left|#1\right\rangle}}
\newcommand{\up}{\ensuremath{\left|\uparrow\right\rangle}}

\newcommand{\down}{\ensuremath{\left|\downarrow\right\rangle}}

\begin{document}

\title{Entanglement and Tunable Spin-Spin Couplings Between \\ Trapped Ions Using Multiple Transverse Modes}
\author{K. Kim$^1$, M.-S. Chang$^1$, R. Islam$^1$, S. Korenblit$^1$, L.-M. Duan$^2$, and C. Monroe$^1$}
\affiliation{$^1$ Joint Quantum Institute, University of Maryland Department of Physics and \\
                    National Institute of Standards and Technology, College Park, MD  20742 \\
             $^{2}$FOCUS Center and MCTP, Department of Physics, University of Michigan,
                     Ann Arbor, Michigan 48109}
\date{\today}
\begin{abstract}
We demonstrate tunable spin-spin couplings between trapped atomic ions, mediated by laser forces on multiple transverse collective modes of motion.  A $\sigma_x \sigma_x$-type Ising interaction is realized between quantum bits stored in the ground hyperfine clock states of \Yb ions.  We demonstrate entangling gates and tailor the spin-spin couplings with two and three trapped ions. The use of closely-spaced transverse modes provides a new class of interactions relevant to quantum computing and simulation with large collections of ions in a single crystal.
\end{abstract}
\maketitle

Trapped atomic ion systems are a leading medium for the generation of quantum entanglement in applications ranging from quantum computing and communication \cite{Nielsen} to quantum simulations \cite{Feynman}.  There have been several recent demonstrations of controlled entanglement of few-ion systems \cite{WinelandBlatt08}, most involving quantum bit (qubit) state-dependent optical forces \cite{CZ,MS}.  Current effort is devoted to the scaling to much larger numbers of trapped ion qubits.  One promising approach entangles small numbers of ions through a coupling to a single mode of collective motion in a single zone, and shuttles individual ions between different trapping zones to propagate the entanglement \cite{QCCD, WinelandBlatt08}.  It may be difficult to entangle more than a few ions through a single mode, owing to potential decoherence or uncontrolled coupling with the many spectator motional modes \cite{NIST}.  However, it may be possible to entangle large numbers of ions through all modes of collective motion, where scalability relies on the high density of motional modes and the relative insensitivity to any particular mode \cite{GZC, Multimode, Uniform}.

In this paper, we demonstrate the tunable interaction between two and three trapped \Yb ions through the simultaneous coupling to multiple closely-spaced transverse modes of motion \cite{TransverseModes, Plenio09, Schaetz09}.  In addition to their high density, transverse modes can be less sensitive to decoherence compared with conventional axial modes \cite{TransverseModes}. We apply an optical spin-dependent force on two ions at a frequency that bisects the two modes of transverse motion in 1-dimension, resulting
in a M\o lmer-S\o rensen $\sigma_x \sigma_x$-type synchronous quantum gate between the ions, and we demonstrate Bell state entanglement fidelities of greater than $95.9\%$ with a gate time of $37.5 \mu$sec.  When the force operates at a frequency far-detuned from the two modes, the motion can be adiabatically eliminated from the Hamiltonian, resulting in a nearly pure spin-spin coupling.  For three ions, we create a GHZ entangled state based on a specific synchronous quantum gate driving all modes.  We finally measure two- and three-spin Ising dynamics and extract the effective spin-spin couplings while varying the detuning of the spin-dependent force.  For larger numbers of ions, this type of control of the form and range of the interaction can be applied to the simulation of complex models of quantum magnetism. \cite{porras04,deng05, Schaetz08}.

We entangle multiple trapped \Yb ions, represented by the $^2S_{1/2}$ hyperfine ``clock" ground states in each ion, abbreviated by the effective spin$-1/2$ states \up and \down, respectively, having frequency splitting $\omega_0/2\pi = 12.643$ GHz \cite{Yb-qubit}. 
We uniformly address the ions with two Raman laser beams in a geometry where the wavevector difference $\delta k$ of the beams points along the transverse $(x)$ direction of trapped ion motion.  Each beam has multiple spectral components that simultaneously drive blue and red motional sideband transitions, following the original scheme of M\o lmer and S\o rensen \cite{MS}, but because the transverse modes of the ions are closely spaced, they all contribute to the interaction.

In general, when noncopropagating laser beams have bichromatic beatnotes at frequencies $\omega_0 \pm \mu$, this can give rise to a spin-dependent force at frequency $\mu$ \cite{Lee05}.  Under the rotating wave approximation
$(\omega_0 \gg \mu)$ and within the Lamb-Dicke limit where $\delta k \langle \hat{x_i} \rangle \ll 1$, with $\hat{x_i}$  the position operator of the $i$th ion, the resulting interaction Hamiltonian is
$H(t) = \hbar \sum_{i} \Omega_i(\delta k\cdot\hat{x}_i)\hat{\sigma}_x^{(i)}$sin$\mu t$ \cite{TransverseModes}.
Here, $\hat{\sigma}_x^{(i)} $ is the Pauli spin flip operator on ion $i$ with Rabi frequency $\Omega_i$ and $\delta k\cdot\hat{x}_i = \sum_{m}\eta_{i,m}(a_m e^{-i\omega_m t} + a^{\dagger}_m e^{i\omega_m t})$ in terms of the normal mode phonon operators $a_m$ and $a^{\dagger}_m$ at frequency $\omega_m$.  The Lamb-Dicke parameters $\eta_{i,m} = b_{i,m}\delta k\sqrt{\hbar/2M\omega_m}$ include the normal mode transformation matrix $b_{i,m}$ of the $i$th ion with the $m$th normal mode \cite{James98}, where $M$ is the mass of a single ion. 

The evolution operator under this Hamiltonian can be written as \cite{Multimode}
\pagebreak
\begin{equation}
U(\tau) = \exp \left[ \sum_{i}\hat{\phi_{i}} \sigma_x^{(i)} 
                   + i\sum_{i,j} \chi_{i,j}(\tau)      \sigma_x^{(i)}\sigma_x^{(j)}\right], 
\label{evolution}
\end{equation}
where 
$\hat{\phi}_i(\tau) = \sum_m[\alpha_{i,m}(\tau) a_m^{\dagger}-\alpha_{i,m}^*(\tau) a_m]$.
The first term in Eq. (\ref{evolution}) represents spin-dependent displacements of the
$m$th motional modes through phase space by an amount  
\begin{equation}
\alpha_{i,m}(\tau) = \frac{-i \eta_{i,m}\Omega_i}{\mu^2 - \omega_m^2} 
                     [\mu - e^{i\omega_m \tau}(\mu \text{cos}\mu \tau - i\omega_m \text{sin}\mu \tau)].
\label{alpha}
\end{equation}
The second term in Eq. (\ref{evolution}) is a spin-spin interaction between ions $i$ and $j$ with coupling 
\begin{eqnarray}
\label{coupling}
\chi_{i,j}(\tau) = &\Omega_i\Omega_j& \sum_m  \frac{\eta_{i,m}\eta_{j,m}}{\mu^2 - \omega_m^2} 
                  \left[ \frac{\mu \text{sin}(\mu-\omega_m)\tau}{\mu-\omega_m} \right. \\ \nonumber
                      &-& \left. \frac{\mu \text{sin}(\mu+\omega_m)\tau}{\mu+\omega_m} 
                      + \frac{\mu sin 2\mu\tau}{2\omega_m} - \omega_m \tau \right].
\end{eqnarray}

We are interested in two regimes where multiple modes of motion contribute to the spin-spin coupling,
taking evolution times $\tau$ to be much longer than the ion osccillation periods $(\omega_m \tau \gg 1)$ \cite{MSgen}.  
In the ``fast" regime, the optical beatnote detuning $\mu$ is close to one or more normal modes and the spins become entangled with the motion through the spin-dependent displacements.  At certain times of the evolution however, $\alpha_{i,m}(\tau) \approx 0$ for all modes $m$ and the motion nearly factorizes, which is useful for synchronous entangling quantum logic gates between the spins \cite{MSgen}. For the special case of $N=2$ ions, both modes in a given direction simultaneously factor when the detuning is set exactly half way between the modes, or at other discrete detunings, where both modes contribute to the effective spin-spin coupling. In the ``slow" regime, the optical beatnote frequency is far from each normal mode compared to that mode's sideband Rabi frequency $(|\mu-\omega_m| \gg \eta_{i,m}\Omega_i)$.  In this case, the phonons are only virtually excited as the displacements become negligible $(\alpha_{i,m} \ll 1)$, and the result is a nearly pure Ising Hamiltonian from the last (secular) term of Eq. (\ref{coupling}):
$H_{eff} = \hbar \sum_{i,j} J_{i,j}\sigma_x^{(i)}\sigma_x^{(j)}$,
where
\begin{equation}
J_{i,j} = -\Omega_i\Omega_j\sum_m\frac{\eta_{i,m}\eta_{j,m}\omega_m}{\mu^2 - \omega_m^2}.
\label{Jij}
\end{equation}
In our experiment, we measure the spin dynamics in both regimes for multiple ions, showing the effect of multiple transverse modes on the interaction.


We trap \Yb ions in a three-layer linear rf-Paul trap \cite{Deslauriers}, with the nearest trap electrode a transverse distance of $100 \mu$m from the ion axis.  An rf potential of amplitude $\sim 200$ V at $39$ MHz is applied to the middle layer, and the top and bottom layers are split into axial segments, where static potentials of order $10-100$ V are applied between the end and center segments for axial confinement. The ions form a linear chain along the trap $z-$axis, and the axial center-of-mass (CM) trap frequency is controlled in the range $\omega_{z}/2\pi = 0.6-1.7$ MHz so that the ions are separated by $2$-$5$ $\mu$m.  In the transverse $x$ direction, the highest frequency normal mode of motion is the CM mode, set to $\omega_{1}/2\pi \sim 4$ MHz.  The second-highest is the ``tilt" mode at frequency  $\omega_{2} = \omega_1\sqrt{1 - \varepsilon^2}$, where the trap anisotropy $\varepsilon = \omega_z/\omega_1 \ll 1$ scales the spacing of the transverse modes.  For three ions, the lowest ``zigzag" mode occurs at frequency $\omega_{3} = \omega_1\sqrt{1 - 2.4\varepsilon^2}$.  
The other transverse $(y)$ modes are sufficiently far away from the $x$ modes (frequencies below $3$ MHz), and the coupling to these modes is suppressed by more than a factor of $10$ by rotating the principal y-axis of motion to be nearly perpendicular to the laser beams \cite{madsen04}.

We direct two Raman laser beams onto the ions to drive spin-dependent forces,
with their wavevector difference aligned along the transverse $x$-axis of ion motion ($\delta k = k\sqrt{2}$, where $k=2\pi/\lambda$). The Raman beams are set to a wavelength of $\lambda = 369.75$ nm, detuned $\sim 500$ GHz to the red of the \Yb\ $^2 S_{1/2} - ^2P_{1/2}$ transition, and have a power of up to $\sim 8$ mW in each beam. The Raman beams are phase modulated at a frequency nearly half of the \Yb hyperfine splitting with a $6.32$ GHz resonant electro-optic modulator, and each of the two Raman beams have independent acousto-optic frequency shifters to select appropriate optical beatnotes to drive Raman transitions in the ions \cite{Lee05}. The Raman beams are focussed to a waist of $\sim 30 \mu$m, and when their beatnote is adjusted to drive the carrier transition at the hyperfine transition $\omega_0$, we observe a carrier Rabi frequency on each ion of up to $\Omega/2\pi \approx 300$ kHz with inhomogeneities between the ions measured to be under $1\%$. 

We first initialize the \Yb ions with 3 ms of Doppler laser cooling on the $^2S_{1/2} - ^2P_{1/2}$ transition at
$369.53$ nm.  We then Raman sideband cool all $m$ modes of transverse motion along $x$ to a mean vibrational indices of $\bar{n}_m<1$ in $\sim 0.5$ ms, well within the Lamb-Dicke limit. Next, the ions are each initialized to the \down state through optical pumping \cite{Yb-qubit}. We then apply the optical spin-dependent force on the ions for a duration $\tau$ by impressing the bichromatic beatnotes at $\omega_0 \pm \mu$ on the Raman laser beams. Afterward, the spin states are measured by directing resonant laser radiation on the $^2S_{1/2}(F=1) - ^2P_{1/2}(F'=0)$ transition, following standard state-dependent fluorescence techniques \cite{Yb-qubit}. We repeat the same experiment $500$ times and average the results. With a single \Yb ion, the effective detection efficiency of the spin state is $97\%$ using a detection time of $0.8$ ms.  For multiple ions, we measure the probability $P_n$ of having $n$ ions in the $\up$ based on the histograms of the collected fluorescence \cite{Sackett}.  
\begin{figure}
\includegraphics[width=1\linewidth]{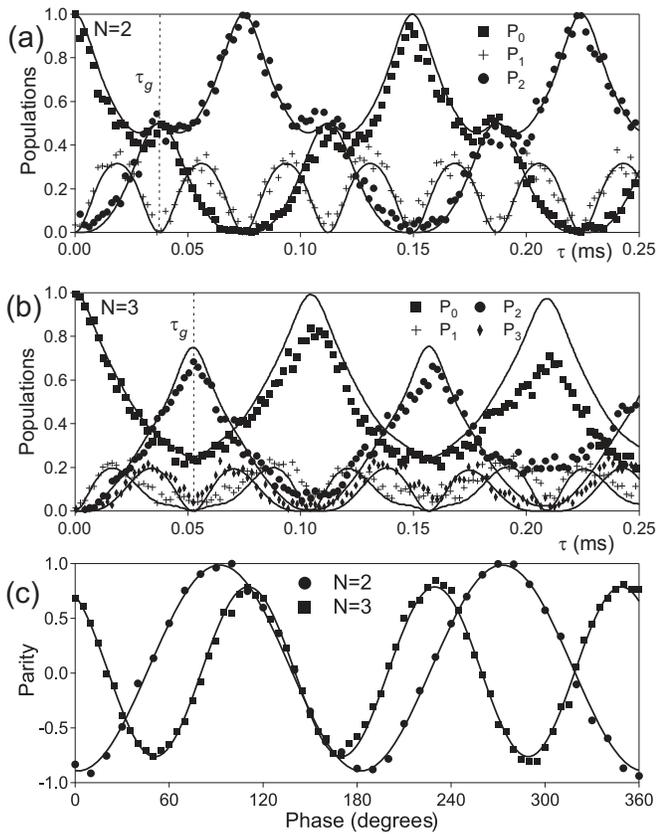}
\caption{(a) Time evolution of the spin populations $P_n$ for two ions subject to a laser force tuned to bisect the transverse CM and tilt sidebands (detuned by $26.7$ kHz from each mode). The ions ideally evolve from $\ket{\downarrow\downarrow}$ to the entangled state 
$\ket{\downarrow \downarrow}+\ket{\uparrow \uparrow}$ after a gate time of $\tau_g = 37.5$ $\mu$s
(dashed line).  (b) Time evolution of the spin populations for three ions subject to a laser force detuned 
$9.4$ kHz blue of the transverse CM mode. The ions ideally evolve from $\ket{\downarrow\downarrow\downarrow}$ 
to the entangled state $\ket{\downarrow \downarrow \downarrow}+\ket{\downarrow \uparrow \uparrow}+\ket{\uparrow \downarrow \uparrow}+\ket{\uparrow \uparrow \downarrow}$ after a gate time of $\tau=53.4$ $\mu$s (dashed line). 
(c) Measured parity of two (circles) and three (squares) ions after the respective gate 
is applied followed by a subsequent global $\pi/2$ rotation of the spins with phase $\phi$.  (For the three-ion
case, an fixed global $\pi/2$ rotation is added before this to create the GHZ state.)  The
contrast of the sinusoidal oscillations with $N\phi$ give the $N-$particle coherence of the state.
The populations at $\tau_g$ in (a) and (b) along with the parity contrast in (c) imply entanglement fidelities of $F_2>95.9\%$ for $N=2$ and $F_3>84\%$ for $N=3$ ions.
The lines in (a) and (b) are fits to Eq. (\ref{evolution}) with no free parameters, and in (c) are sinusoidal 
fits to extract contrast.}
\label{gate}
\end{figure}

For entanglement of $N=2$ ions through multiple modes, we set $\omega_{z}/2\pi = 0.616$ MHz,  
$\omega_1/2\pi = 3.5838$ MHz, and $\omega_2/2\pi = 3.5305$ MHz.  Fig. \ref{gate}(a) shows the evolution of the spin populations $P_0$ and $P_2$ when the Raman coupling is applied with an optical beatnote detuning $\mu/2\pi = (\omega_1+\omega_2)/4\pi = 3.5571$ MHz that bisects the CM and tilt mode sidebands, 
$26.7$ kHz away from each sideband. We adjust the Raman laser power so that after the gate time 
$\tau_g = 2\pi/(\omega_1-\mu) = 37.5\ \mu$s, 
the spin-spin coupling from Eq. (\ref{coupling}) is $\chi_{1,2} = \pi/4$, or 
$\eta\Omega = (\omega_1-\mu)/2\sqrt{2}$, where the magnitudes of the Lamb-Dicke parameters 
for either ion and either mode are all approximately $\eta = 0.049$.  
At time $\tau_g$, we measure $P_0+P_N = 97.8(1.6)\%$.  To characterize the coherence of the entangled state, we apply the same gate followed by a global Raman carrier $\pi/2$-pulse with phase $\phi$ relative to the gate, and measure the
parity $\sum_n (-1)^n P'_n$ of the resulting state, where $P'_n$ are the new populations following
the rotation.  The contast of the parity oscillating as sin$N\phi$ 
[circles in Fig. \ref{gate}(c)] of $94(1)\%$ for $N=2$ indicates the fidelity of the two-ion entangled state 
$F_2 >96(2)\%$ \cite{Sackett}. After $15$ consecutive gate operations ($562\ \mu$s), the entanglement fidelity degrades to 74(4)$\%$, implying a single-gate fidelity of $98\%$ exclusive of state preparation and detection errors.  
For $N=3$ ions, we set the trap frequencies to $\omega_{z}/2\pi = 1.484$ \MHz, $\omega_{1}/2 \pi =  3.952$ \MHz, $\omega_{2}/2\pi = 3.663$ MHz, and  $\omega_{3}/2\pi = 3.215$ MHz, and create a GHZ entangled state
by tuning the Raman beatnote blue of the CM sideband by $(\mu-\omega_1)/2\pi = 9.4$ kHz. While the other modes do not contribute nearly as much as the CM, they must be considered for the appropriate gate time and Raman power for the operation.  Here, at time $\tau_g = 2\pi/(\omega_1-\mu) = 53.4\ \mu$s, we expect $\chi_{i,j} = \pi/4$ for all spin pairs when the Raman power is set such that $\eta'\Omega \approx (\omega_1-\mu)/1.95$, where $\eta' = 0.0380$ is
the 3-ion CM Lamb-Dicke parameter.  Fig. \ref{gate}(b) shows the time evolution of the state, and after the gate time $\tau_g$ a simple $\pi/2$ rotation of all the spins ideally creates a GHZ state 
$\ket{\uparrow \uparrow \uparrow}+\ket{\downarrow \downarrow \downarrow}$. By applying a subsequent carrier $\pi$/2 pulse with variable phase as above and measuring the constrast of the parity oscillations and populations 
[Fig \ref{gate}(c)], we arrive at a three-ion GHZ entanglement fidelity of $84(3)\%$ [$P_0+P_N = 91(2)\%$, parity contrast $=77(1)\%$].  The main sources of gate infidelity and decoherence are classical laser power fluctuations and spontaneous emission from the Raman beams.  Additional terms not considered in the model Hamiltonian, including slight differential AC-Stark shifts, off-resonant transitions on the carrier, and higher order Lamb-Dicke effects are expected to contribute less than $2\%$ to fidelity errors.

Next we set the bichromatic detuning $\mu$ to be far from all motional modes so that the evolution is dominated by pure spin-spin couplings. In this case, we measure $P_0(\tau)$ as well as the average number of ions in state \up $F(\tau) = \sum_n n P_n(\tau)$ after various durations $\tau$ of the bichromatic force. We can then extract the spectrum of couplings $J_{i,j}$ by observing the secular Fourier components in the time evolution 
[Eq. (\ref{evolution})].  

\begin{figure}
\includegraphics[width=1\linewidth]{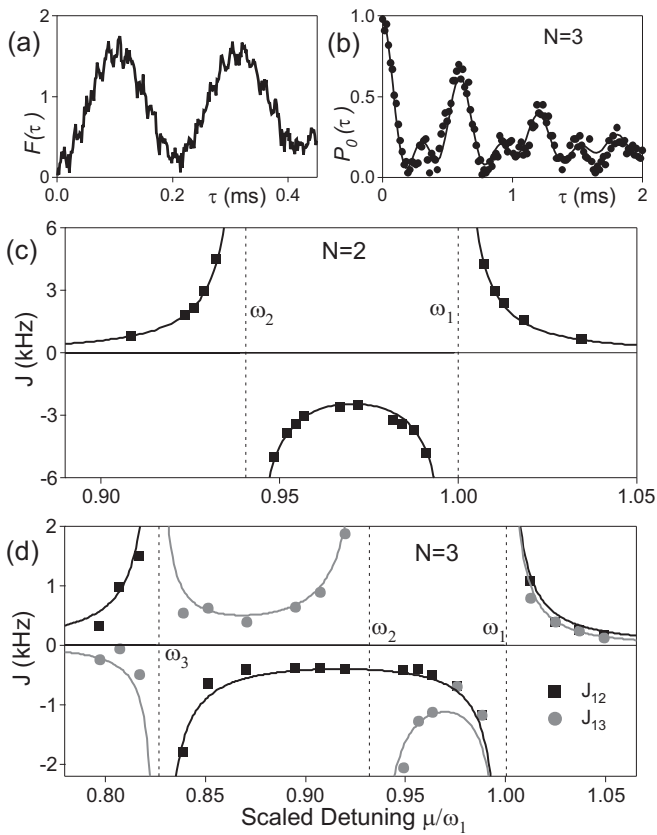}
\caption{(a) Time evolution of the average number of ions in \up for $N=2$ ions under the bichromatic force in the far-detuned limit, showing the secular oscillation of the Ising spin-spin coupling at $4.8$ kHz accompanied by small fast modulations from motional excitation at $\sim56$ kHz. (b) Time evolution of the probability $P_0$ of all spins \down, with $N=3$ ions under the bichromatic force in the far-detuned limit. Here, the fast motional oscillations are negligible, but the two couplings $J_{1,2}=J_{2,3}$ and $J_{1,3}$ are clearly visible. Solid line is a fit to the secular terms in Eq. (\ref{coupling}) with an empirical exponential decay. (c-d) Measured couplings $J_{i,j}$ for two (c) and three (d) ions (points) as a function of detuning $\mu$ overlaid with theory (lines) from Eq. (\ref{Jij}) with no free fit parameters (the overall sign in (d) is set from theory). The normal modes positions are indicated by the dashed lines.}
\label{FigJ2J3}
\end{figure}

For N=2, the observed spin evolution is a simple sinusoid shown in Fig. \ref{FigJ2J3}(a), accompanied by low-contrast fast undulations from the residual motional couplings [first three terms of Eq. (\ref{coupling})]. We plot the measured couplings $J_{1,2}$ as the detuning $\mu$ is varied in Fig. \ref{FigJ2J3}(c), along with that expected from Eq. (\ref{Jij}) with no adjustable parameters. We determine the sign of the coupling by preceding the spin-dependent force with a carrier $\pi/2$ pulse and measuring the phase of the spin evolution.

For $N=3$ ions, there is competition between the nearest neighbor couplings $J_{1,2} = J_{2,3}$ and the next-nearest-neighbor coupling $J_{1,3}$, as seen in the time evolution of $P_0$ in Fig. \ref{FigJ2J3}(b).  By scanning the detuning $\mu$, again we see from Fig. \ref{FigJ2J3}(d) that the form of the couplings can be 
controlled.  For example, in regimes where the next-nearest-neighbors are anti-ferromagnetic ($J_{1,3}>0$), the system is expected to exhibit magnetic frustration. In the limit of very large detuning $(|\mu-\omega_i| \gg \omega_1 - \omega_3)$, we expect a dipolar decay of the couplings $J_{1,3} = J_{1,2}/8$ from Eq. (\ref{Jij}) 
for the equally-spaced ions.

In summary, the experiment demonstrates how to tailor spin-spin Hamiltonians in trapped ion crystals by controlling the detuning of a spin-dependent force in the region of closely-spaced transverse motional modes, and this approach should be scalable to much larger numbers of ions.  
One future direction is to add an effective magnetic field (e.g., $B\sum_i\sigma^{(i)}_y$) transverse to the Ising couplings, and characterize nontrivial phases of this system for larger numbers of ions 
\cite{porras04, Schaetz08}.  

\begin{acknowledgments}
We acknowledge useful discussions with J. Freericks and G.-D. Lin.
This work is supported by the DARPA Optical Lattice Emulator Program, IARPA under ARO contract, the NSF Physics at the Information Frontier Program, and the NSF Physics Frontier Center at JQI.

\end{acknowledgments}

\bibliography{TransverseMS}

\end{document}